\documentclass[FBSedit,FBSmath,ecsub]{FBSsuppl}
\usepackage{amsmath}
\usepackage{amsfonts}
\usepackage{amssymb}

\usepackage{graphicx}

\title{Three-Body Dynamics in One Dimension}

\author{T. Melde \instnr{1}, L. Canton\instnr{1,2}, J.P. Svenne \instnr{3}}
\instlist{ 
Dipartimento di Fisica, Universit\`a di Padova, 
Via F. Marzolo 8, Padova, Italy, I-35131
\and Istituto Nazionale di Fisica Nucleare, Sezione di Padova,
Via F. Marzolo 8, Padova, Italy, I-35131
\and
Winnipeg Institute for Theoretical Physics and 
Department of Physics and Astronomy,
University of Manitoba, 
Winnipeg, MB, Canada R3T 2N2
}

\begin{document}
\maketitle
\begin{abstract}
    We discuss the general three-particle quantum scattering 
problem, for motion restricted to the full line. 
Specifically, we formulate the three-body problem in one dimension
in terms of the (Faddeev-type) integral equation approach.
As a first application, we develop a spinless, one-dimensional (1-D) model
that mimics three-nucleon dynamics in one dimension. In addition, we 
investigate the effects of pionic 3NF diagrams.
\end{abstract}
\section{Introduction}
The quantum mechanics of one-dimensional systems has been extensively 
investigated for a variety of purposes~\cite{BGKW85}. Quantum scattering 
in one dimension
has also been considered because of its value in pedagogical
works~\cite{BLA01}, since it retains sufficient complexity to
embody many of the physical concepts and behaviours that occur in the
more complex three-dimensional processes, but without the many technical
aspects required for dealing with quantum scattering in higher 
dimensions.
Here, we are 
interested in the formulation of the quantum three-body theory on the
full,  $(-\infty,+\infty)$ line, based on a system of coupled integral
equations for the three-body transition operators~\cite{Mel01}  (AGS
equations~\cite{AGS}). 
We illustrate in
this paper an approach that harmonizes the typical two-channel structure
of the ``tunneling'' problems, due to the presence of the transmission
and reflection processes, with the cluster structure of the three-body
problem, thus leading to an overall reduced S-matrix
for the two-fragment processes which is
$6\times 6$~\cite{MCS02}. The formulation presented herein, with straightforward
modifications, is suited also to describe the tunneling (or
potential-barrier) effects due to the scattering of a two-body
one-dimensional cluster impinging on an external barrier potential.
\section{Theory}
The Schr\"odinger equation for a one-dimensional system on 
the whole line,
\begin{equation}
-\frac{d^2\Psi_k \left( x\right)}{dx^2}
+U\left( x\right)\Psi_k \left( x\right)
=k^2\Psi_k \left( x\right)
\label{eq:Schroe}
\end{equation}
defines the transmission-reflection problem for positive energies
and has two independent solutions with
incident waves from the left and right, respectively.
The asymptotic form
for the wave function, with incidence from the left is 
\begin{equation}
\Psi_k^L\left( x\right)\rightarrow
\begin{cases}
e^{i kx}+R_L \left( E\right)e^{-i kx},
& \text{for $x \rightarrow -\infty$} \\
T_L \left( E\right)e^{i kx},
& \text{for $x \rightarrow +\infty$}
\end{cases}
\label{eq:Schroe-sol}
\end{equation}
where $T_L,R_L$ are the transmission and reflection coefficients, with 
a similar expression for incidence from the right.
The S-matrix is given according to the definition
\begin{equation}
S \left( E\right)=
\begin{pmatrix}
T_L  \left( E\right)& R_R  \left( E\right)\\
R_L  \left( E\right)& T_R  \left( E\right)\\
\end{pmatrix}
\label{eq:Smatrix}
\end{equation}
which reduces to the identity matrix when the potential
goes to zero.
Off the energy-shell the t-matrix is
introduced according to
\begin{equation}
t\left( E\right)\left|\Phi_q\right>=V\left|\Psi_q\right>
\end{equation}
where
$
\left|\Phi_q\right>
$
is the free (plane-wave) solution.
The S-matrix in respect to the on-shell t-matrix is
\begin{equation}
S\left( E\right)=
\begin{pmatrix}
	1 & 0\\
	0 & 1\\
    \end{pmatrix}
-\frac{i m}{\hbar^2 k}
\begin{pmatrix}
t\left(k,k;E\right)  & t\left(k,-k;E\right)\\
t\left(-k,k;E\right) & t\left(-k,-k;E\right)
\end{pmatrix}
\label{eq:STconn}
\end{equation}
where $t(k,k')$ are the on-energy-shell matrix elements
of the t-matrix.
In the one-dimensional system, on the energy shell
for a given energy $E$, there exist exactly two possible momenta
$
k=k_\pm=\pm \frac{\sqrt{2mE}}{\hbar}
$. In the three-body system it is possible to define a S-matrix for 
two-fragment scattering in a similar way.

Generally, a system of three particles is described by the Hamiltonian
\begin{equation}
H=H_0+\sum_a{V_a}
\end{equation}
and it is well established to introduce so called channel states, which 
are eigenstates of the channel Hamiltonians
$
H_a=H_0+V_a
$
with $a=1,2,3$. 
In the one-dimensional system the channel states can
be described by the following wave functions
\begin{equation}
\Phi_{\pm a}=\Phi_{\pm q_a}\left(x_a,y_a\right)
=e^{\pm i q_ay_a}\phi_a\left(x_a\right)
\end{equation}
where $\phi_a\left(x_a\right)$ denotes the properly
normalized bound-state wave function of the two-particle
fragment and $e^{\pm i q_ay_a}$ denotes
the relative motion of the particle $a$ with respect to the
c.m. of the two-particle fragment $\left(bc\right)$.
The asymptotic 1-D plane wave incoming from the left can have three
different transmitted two-cluster waves, which leads to the asymptotic form
\begin{equation}
\Psi_{a}^L\left(x,y\right)\rightarrow
\begin{cases}
\Phi_{+b}\delta_{ab}+R^L_{ab}\left( E\right)\Phi_{-b},
& \text{for $  y_b\rightarrow -\infty$}\\
T^L_{ab}\left( E\right)\Phi_{+b},
& \text{for $ y_b\rightarrow +\infty$}
\end{cases}
\end{equation}
and the reduced S-matrix for two-fragment scattering is
\begin{equation}
 { \tilde S_{ab}}\left(E\right)=
\begin{pmatrix}
T^L_{ab}\left(E\right) & R^R_{ab}\left(E\right) \\
R^L_{ab}\left(E\right) & T^R_{ab}\left(E\right)
\end{pmatrix} 
\label{eq:TR3space}
\end{equation}
This definition produces a {$6\times 6$ }
scattering matrix connecting all
six possible ``two-fragment asymptotic channels''.
We have shown~\cite{MCS02}, that it is possible to define the 
two-fragment S-matrix in one dimension in the following way
\begin{multline}
    { \tilde S_{ab}}\left(E\right)  
    =
    \left\langle \phi_{\pm a}\right|\tilde S_{ab}\left|\phi_{\pm b}\right\rangle
    \\=
    \delta_{ab}
    \begin{pmatrix}
	1 & 0\\
	0 & 1\\
    \end{pmatrix}
    -2\pi \frac{i M}{\hbar^2\left|q_b\right|}
    \begin{pmatrix}
	t_{ab}\left(+\left|q_a\right|,+\left|q_b\right|;E\right) &
	t_{ab}\left(+\left|q_a\right|,-\left|q_b\right|;E\right) \\
	t_{ab}\left(-\left|q_a\right|,+\left|q_b\right|;E\right) &
	t_{ab}\left(-\left|q_a\right|,-\left|q_b\right|;E\right)
    \end{pmatrix} 
\end{multline}
where $t_{ab}$ are related to the AGS channel-transition operators, 
$U_{ab}$.
It easily seen that the structure of this equation is very similar to 
the corresponding one in the two-body system, with the difference 
that here the matrix space is expanded to account for the 2+1 
partitions. 
\section{Results}
\begin{figure}[hbt]
\centerline
{
\includegraphics[width=4cm]{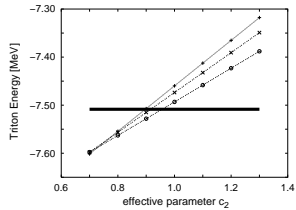}
}
 \caption{Dependence of the binding energy on the parameter
 $c_2$ for all three potentials including
 the diagonal correction terms.
 The thick line denotes the triton energy without any correction terms
 for all three potentials. }
\label{fig:para}
\end{figure}

As first application, we develop a one-dimensional spinless model that
mimics three-nucleon dynamics in one dimension.
In particular, we investigate numerically the effects of 
irreducible  ``pionic'' corrections~\cite{Can98,CMS01} in the 
one-dimensional three-body bound state by solving the homogeneous version 
of the AGS equation.
Three two-body potentials that give the same  2N binding at 
$-2.225$MeV but with different  ``stiffness'' of the repulsion terms 
have been considered. 
\begin{table}[hbt]
\begin{tabular}{ccccc}
\hline
         & {2NF} 
	 & {2NF+OPE3}
	 & {2NF+TPE3 }
	 & {2NF+OPE3+TPE3} \\
\hline
  {$V_5$ }  & -7.722 & -7.708 & -8.733 & -8.718 \\
\hline
 {$V_{10}$ }& -7.684 & -7.662 & -8.670 & -8.650 \\
\hline
 {$V_{20}$ }& -7.668 & -7.642 & -8.631 & -8.608 \\
\hline
    \end{tabular}
\caption{
 `Triton' energies in MeV for the
 potentials with different ``stiffness'', including
 different 3NF corrections.}
 \label{tbl:Econv}
\end{table}
In table~\ref{tbl:Econv} we show the effect of one-pion (OPE3) and 
two-pion exchange (TPE3) three-nucleon force (3NF) type corrections. 
For the details of the explicit form of these correction terms we 
refer to the literature~\cite{CMS01,MCS02}. It is seen that unlike the 
TPE, the OPE corrections have only a small effect on the `triton'
binding energies. In figure~\ref{fig:para} we show the effect of an 
effective parameter required in the OPE 3NF~\cite{CS01,CSH02} (set equal to 
one in table~\ref{tbl:Econv}).
It is seen that for values not equal to one, the effects on the 
`triton' binding energy can be non-negliglible and the size 
of the effect depends directly on the `stiffness' of the potential.

In conclusion we have developed an integral formulation of the 
one-dimensional scattering problem on the full line. As first 
application we tested the effects of different 3NF-type correction 
terms on the triton binding energy. It would be interesting to extend 
the investigations of the 2+1 fragment 1D-system to the scattering 
regimes.
\begin{acknowledge}
    We acknowledge support from the Italian MURST-PRIN Project ``Fisica 
    Teorica del Nucleo e dei Sistemi a Pi\`u Corpi''. 
    J.P.S. acknowledges support from NSERC, Canada. 
\end{acknowledge}
\end{document}